\documentstyle[array,multicol,aps,rmp,psfig]{revtex}
\begin{document}

\title{\Large \bf Ocular dominance patterns in mammalian visual cortex: 
A wire length minimization approach}
\author{Dmitri B. Chklovskii and Alexei A. Koulakov}
\address{Sloan Center for Theoretical Neurobiology, The Salk Institute, La Jolla, CA 92037}

\date{June 14, 1999}
\draft
\tighten
\preprint{Submitted to {\it Journal of Neuroscience}}
\maketitle

\begin{abstract}
We propose a theory for ocular dominance (OD) patterns in mammalian
primary visual cortex. This theory is based on the premise that OD
pattern is an adaptation to minimize the length of intra-cortical
wiring. Thus we can understand the existing OD patterns by solving a
wire length minimization problem. We divide all the neurons into two
classes: left-eye dominated and right-eye dominated. We find that 
segregation of neurons into monocular regions reduces wire length if 
the number of connections with the neurons of the same class differs 
from that with the other class. The shape of the regions depends on the
relative fraction of neurons in the two classes. If the numbers are
close we find that the optimal OD pattern consists of interdigitating
stripes. If one class is less numerous than the other, the optimal OD
pattern consists of patches of the first class neurons in the sea of
the other class neurons. We predict the transition from stripes
to patches when the fraction of neurons dominated by the ipsilateral 
eye is about $40\%$. This prediction agrees with the data in macaque and Cebus
monkeys. This theory can be applied to other binary cortical systems.
\end{abstract}

\begin{multicols}{2}

\section{INTRODUCTION}

In the primary visual area (V1) of many mammals, most neurons respond
to the stimulation of the two eyes unevenly: they are either left-eye
or right-eye dominated. In some species, left-/right-eye dominated
neurons are uniformly intermixed in space. In others, left-/right-eye 
dominated neurons are segregated resulting in a system of alternating 
monocular regions. This system is known as the ocular dominance (OD) 
pattern~\cite{WHL}.

Most theorists interested in the OD pattern~\cite{Erwin}, \cite{SwinRev} have
been modeling its development.  They succeeded in generating OD
patterns of realistic appearance. However, several {\em why} rather
than {\em how} questions remained unanswered. Why, from the functional
point of view, do the OD patterns exist?  Why do some mammalian
species have OD patterns while others do not? Why do monocular regions
have different appearances (stripes as opposed to patches) between
different species and even between different parts of V1 within the
same animal?

Mitchison, 1991, suggested an answer to the first question using the
wiring economy principle~\cite{Cajal}, \cite{AK}, \cite{Cowey},
\cite{Chern}, \cite{Young}, \cite{CS}. The idea
is that the evolutionary pressure to keep the brain volume to a
minimum requires making the wiring (axons and dendrites) as short as
possible, while maintaining function. In general, the function of a
cortical circuit specifies the connections between neurons (wiring
rules). Therefore the problem presented by the wiring economy
principle is to find, for given wiring rules, the spatial layout of
neurons that minimizes wire length. Then we can understand the
existing layout of neurons as a solution to the wire length
minimization problem.

We adopt the wiring economy principle and address the above questions
by formulating and solving a wire length minimization problem. Because
of the columnar organization of the cortex~\cite{Mou} we consider a
two-dimensional neuronal layer of uniform density. The number of the
left-eye dominated neurons is a fraction $f_L$ of the total number,
and $f_R$ is a fraction of right-eye dominated neurons ($f_L+f_R=1$).

We consider only intra-cortical connections because they constitute
the majority of gray matter wiring~\cite{LVG}, \cite{PP}, \cite{Ahmed} allowing us to
neglect the thalamic afferents and other extra-cortical
projections. We assume that each neuron receives synapses from $N_{s}$
neurons dominated by the same eye and from $N_{o}$ neurons dominated
by the opposite eye. In other words, because synapses are unidirectional 
the resulting wiring rules require each neuron to get unidirectional 
connections from $N_{s}$ neurons dominated by the same eye and from $N_{o}$
neurons dominated by the opposite eye.

\begin{figure}
\centerline{
\psfig{file=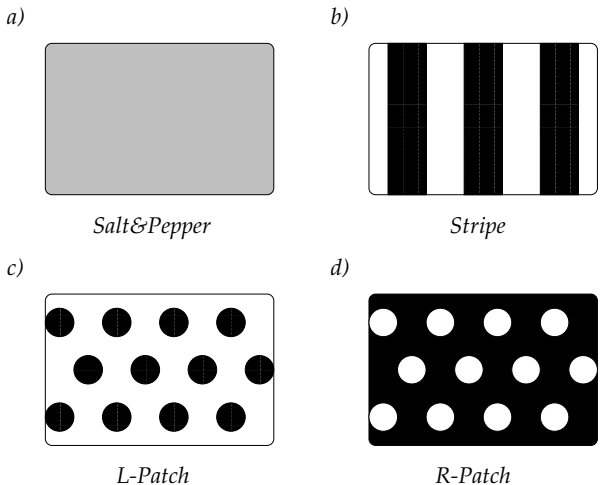,width=3.1in}
}
\vspace{0.1in} 
\setlength{\columnwidth}{3.4in}
\caption{
Different appearances of the ocular dominance
pattern. Left-eye dominated neurons are black while right-eye dominated
neurons are white.(a) {\em Salt and Pepper} phase, uniformly intermixed
left/right neurons.(b) {\em Stripe} phase, alternating monocular
stripes.(c) {\em L-Patch} phase, circular left-eye islands in the
right-eye sea. (d) {\em R-Patch} phase, circular right-eye islands in
the left-eye sea.
\label{phases}
}
\vspace{-0.1in}
\end{figure}

Given these wiring rules we look for an optimal layout of neurons
which minimizes the total length of connections.  Depending on 
the values of $N_s$, $N_o$, and $f_L$, optimal layout
belongs to the one of the four phases shown in Fig.\ref{phases} where
left-eye dominated neurons are shown in black and right-eye dominated
neurons - in white. In the {\em Salt and Pepper} phase left-eye and
right-eye dominated neurons are uniformly intermixed,
Fig.\ref{phases}a.  The {\em Stripe} phase consists of alternating
monocular stripes of neurons dominated by either eye,
Fig.\ref{phases}b. The {\em L-Patch} consists of the patches of the
left-eye dominated neurons surrounded by the right-eye dominated
neurons, Fig.\ref{phases}c. The {\em R-Patch} consists of the patches
of the right-eye dominated neurons surrounded by the left-eye
dominated neurons, Fig.\ref{phases}d.

Our approach differs from Mitchison's in that we drop the retinotopy
requirement, that is our wiring rules do not take into account
receptive field positions. This simplification is supported by the
existence of the receptive field scatter~\cite{HubWies74}, random
variation in the receptive field position between adjacent
neurons. For example, in a macaque retinotopy exists only on the
scales greater than $\approx 1$mm~\cite{HubWies74}, which exceeds the
typical size of monocular regions. By disentangling the retinotopy
from the OD problem we simplify it enough to map out a phase diagram.

In the Discussion we compare our predictions with the data from macaque 
and Cebus monkeys and find good agreement. Also, we discuss
simplifying assumptions made in the paper and possible ways to extend
the theory.

%
%


\section{RESULTS}

\begin{figure}
\centerline{
\psfig{file=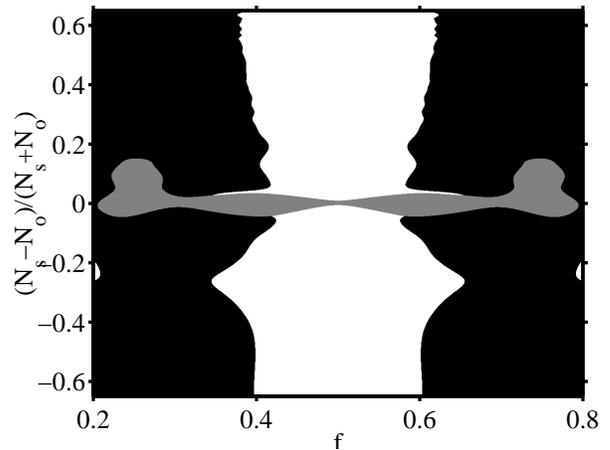,width=3.1in}
}
\vspace{0.1in} 
\setlength{\columnwidth}{3.4in}
\caption{Ocular dominance phase diagram calculated in the
lattice model. Optimal phases
are shown as a function of the relative number of the same-eye to
other-eye connections $N_s/N_o$ and a fraction of left-eye neurons
$f$. Range of the {\em Stripe} phase optimality is shown in
black, {\em Patch} phase - in white, {\em Salt and Pepper} - in grey.
\label{ph_diagram}
}
\vspace{-0.1in}
\end{figure}

We present the central results of the paper on a phase diagram,
Fig.\ref{ph_diagram}, showing optimal phases for various ratios of
same-eye to other-eye connections $N_s/N_o$ and fractions of left-eye
neurons $f_L$. If the numbers of same-eye and other-eye connections
are equal, $N_{s}/N_{o}=1$ then {\em Salt and pepper} phase is
optimal. Otherwise, if $N_{s}/N_{o}\neq 1$ the wirelength is minimized
by an OD pattern consisting of alternating monocular
regions. The shape of these regions depends on the relative fraction
of the left-eye dominated neurons, $f_L$. When the numbers of neurons
dominated by each eye are close, $f_L\approx f_R$, the {\em Stripe}
phase is optimal. When the fraction of left-eye (right-eye) dominated
neurons drops below a critical value $f_c\approx 0.4$ the {\em
L-Patch} ({\em R-Patch}) phase becomes optimal. Our predictions of the
critical value agree with the data from macaque and Cebus monkeys.

%
%
In the following section we formulate a wiring problem on a lattice. For 
small $N_s$ and
$N_o$ we solve it analytically while for large $N_s$ and $N_o$ we solve it
numerically. Results are shown in Fig.\ref{ph_diagram}.
Next, we introduce a continuous formulation of the problem. We prove that
{\em Salt and Pepper} is an optimal layout when
$N_s=N_o$. Then we show that for $N_s\neq N_o$ segregation of neurons
according to their OD reduces wire length. We calculate in perturbation
theory the wire length for {\em Stripe} and {\em Patch} phases and 
find the range of parameters for the optimality of each phase. 
Perturbation theory provides an analytical treatment of neuronal 
clustering so common throughout the nervous system. The
calculated phase diagram is similar to that obtained in the lattice 
model.

\subsection{Lattice model}

Although the arrangement of neurons in cerebral cortex is anything but
grid-like we can understand many features of the neuronal layout by
studying lattice models. These models compensate in clarity and
computability what they lack in realism. Of course, we need to make
sure that the results are independent of the particular choices of
lattice parameters (for example the number of nearest neighbors).

We consider arranging a large number of neurons on a two-dimensional
square lattice. Each site must be occupied by either left-eye or
right-eye dominated neuron. The number of the left-eye dominated neurons 
is a fraction $f_L$ of the total number of neurons and $f_R$ is a
fraction of right-eye dominated neurons. The problem is to find a 
layout which minimizes the total
length of wiring specified by the following rule. Each left-eye
neuron has unidirectional connections with $N_{s}$ left-eye
neurons and with $N_{o}$ right-eye neurons. Each right-eye
neuron has unidirectional connections with $N_{s}$ right-eye
neurons and with $N_{o}$ left-eye neurons. Unidirectionality of
connections means that connecting neuron A to neuron B, does not
necessarily imply that neuron B connects to neuron A. The motivation
for this rule comes from the unidirectional properties of synapses in
the brain.

Because we attempt to minimize wire length we assume that for a given
layout the connections are established optimally. Thus the problem is
reduced to comparing optimal wiring for various layouts. Therefore, we
will assume that each neuron makes the shortest possible connections
satisfying wiring rules.

{\it Small numbers of connections per neuron.}
We start by finding optimal layouts for three illustrative examples of 
wiring rules with small numbers of connections, $N_s$ and $N_o$. We 
caution the reader that because of the small numbers of connections
phase assignments may seem arbitrary. These examples are
chosen to {\em illustrate} our main results which will be confirmed both in
the lattice model with large $N_s$ and $N_o$ later in this section and in 
the continuous model (section IIB).

\begin{figure}
\centerline{
\psfig{file=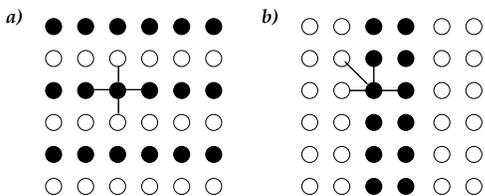,width=2.5in}
}
\vspace{0.1in} 
\setlength{\columnwidth}{3.4in}
\caption{ Ocular dominance patterns for $f_L=1/2$ and
$N_s=N_o=2$.  (a) A realization of the {\em Salt and Pepper} phase
gives minimal wire length.  (b) A realization of the {\em Stripe}
phase is suboptimal.
\label{lat1}
}
\vspace{-0.1in}
\end{figure}

For the first two examples we set equal numbers of left and right
neurons, $f_L=f_R=1/2$.  In the first example each neuron connects
with equal numbers of the same-eye and other-eye neurons,
$N_s=N_o=2$. Then the optimal layout is the system of alternating rows
of left/right neurons, Fig.\ref{lat1}a. This layout is a realization
of the {\em Salt and Pepper} phase, Fig.\ref{phases}a, because each
neuron has an equal number of left and right neurons among its nearest
neighbors. To calculate the length of connections per neuron, $l$, we
notice that in this layout all neurons have the same pattern of
connections. By considering one of them, Fig.\ref{lat1}a, we find that
$l=4$. This layout is optimal because each neuron makes all of its
connections with nearest neighbors.

%
%

A suboptimal layout for the same wiring rules is illustrated by a
realization of the {\em Stripe} phase, Fig.\ref{lat1}b. In this layout
each neuron has the same pattern of connections up to a mirror
reflection. By considering one of them, Fig.\ref{lat1}b, we find
$l=3+\sqrt{2}\approx 4.41$, greater than $l=4$ for the {\em
Salt and Pepper} phase. Here each neuron has among its nearest
neighbors only one other-eye neuron, while the wiring rules require
connecting with two other-eye neurons. A connection to the next
nearest neighbor is longer making the layout suboptimal. We confirm
the optimality of the {\em Salt and Pepper} phase for $N_s=N_o$ both
numerically for large $N_s$, $N_o$ and analytically.

\begin{figure}
\centerline{
\psfig{file=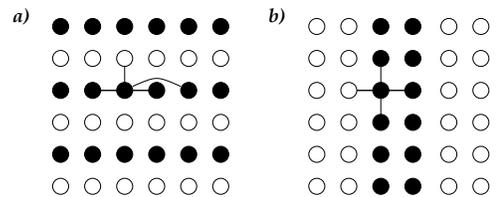,width=2.5in}
}
\vspace{0.1in} 
\setlength{\columnwidth}{3.4in}
\caption{ Ocular dominance patterns for $f_L=1/2$ and
$N_s=3$, $N_o=1$.  (a) A realization of the {\em Salt and Pepper} is
suboptimal.  (b) A realization of the {\em Stripe} phase gives minimal
wire length.
\label{lat2}
}
\vspace{-0.2in}
\end{figure}

In the second example each neuron connects with more same-eye than
other-eye neurons: $N_s=3$, $N_o=1$. Then a realization of the {\em
Salt and Pepper} phase, Fig.\ref{lat2}a is not optimal anymore. The
length of connections per neuron is $l\approx 4.41$, while the {\em
Stripe} phase, Fig.\ref{lat2}b gives $l=4$. The {\em Salt and Pepper}
phase loses in wiring efficiency because there are not enough same-eye
neurons among nearest neighbors and connections with the next nearest
neighbors are needed.  The {\em Stripe} phase, Fig.\ref{lat2}b
rectifies this inefficiency by having each neuron make connections
only with nearest neighbors. Thus, clustering of same-eye neurons is
advantageous if each neuron connects more with the same-eye than with
the other-eye neurons.

%
%


%
%
\begin{figure}
\centerline{
\psfig{file=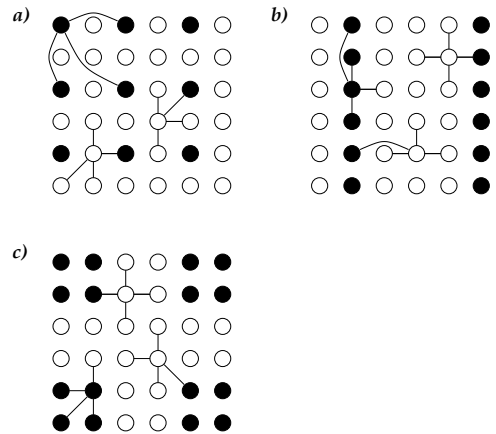,width=2.5in}
}
\vspace{0.1in} 
\setlength{\columnwidth}{3.4in}
\caption{ Ocular dominance patterns for $f_R=3/4$ and
$N_s=3$, $N_o=1$.  Realizations of the {\em Salt and Pepper} (a) the
{\em Stripe} (b) are suboptimal.  (c) A realization of the {\em
L-Patch} phase gives minimal wire length.
\label{lat3}
}
\vspace{-0.2in}
\end{figure}

In the third example we use the same wiring rules ($N_s=3$, $N_o=1$)
but take different numbers of left/right neurons, $f_L=1/4$,
$f_R=3/4$. The realizations of the {\em Salt and Pepper} phase is
shown in Fig.\ref{lat3}a and of the {\em Stripe} phase in
Fig.\ref{lat3}b. In these layouts, different neurons have different
patterns of connections. To find the wiring length per neuron we
average over different patterns and find for the {\em Salt and Pepper}
phase $l\approx 5.02$ and for the {\em Stripe} phase $l=4.5$. A more
efficient layout is the {\em L-Patch} phase, Fig.\ref{lat3}c, where
$l\approx 4.21$. Although we cannot prove that the {\em L-Patch}
phase is optimal, this seems likely. Thus, the optimal shape of
monocular regions depends on the relative numbers of left/right
neurons.


{\it Large numbers of connections per neuron.}
Lattice models with small numbers of connections per neuron yield
quick results good for illustration purposes. However, they are
difficult to generalize to the wiring rules with large numbers of
connections more appropriate for cortical circuits where each neuron
connects with $\approx10^4$ neurons.  For example, attributing the
layouts in Fig.\ref{lat1}a,\ref{lat2}a to the {\em Salt and Pepper}
phase rather than the {\em Stripe} phase may seem arbitrary. Therefore
we study lattice problems with large numbers of connections per
neuron.

When the number of connections per neuron is much greater than the
number of nearest neighbors, the effect of the discreteness of the
lattice on the results is negligible. In particular, for a given
fraction $f_L$ and ratio $N_s/N_o$ only the periodicity of the optimal
layout depends on the $N_o$. Thus the solution of the wire length
minimization problem for one value of $N_o$ can be generalized to
other problems with the same $f_L$ and $N_s/N_o$.

We solve the wire length minimization problem using the following numerical
algorithm. We fix the values of $f_L$ and $N_s/N_o$. We consider
neuronal layouts belonging to several phases: {\em Salt and Pepper},
{\em Stripe}, {\em L-(R-)Patch} (both triangular and square lattice),
{\em Checkerboard} (only for $f_L=1/2$). For each phase we find the
optimal period which minimizes wirelegth. Then we compare wire length
in the optimal layouts of different phases. We plot the optimal phases
for various values of $f_L$ and $N_s/N_o$ on the phase diagram,
Fig.\ref{ph_diagram}. These results were discussed above.

\subsection{Continuous model}

In this section we study the limit when $N_s$ and $N_o$ are very large. 
Instead of considering each receiving connections neuron separately
it makes sense to treat them as a mixture of two "liquids",
the left- and right- eye ones, having continuous in space densities.
Segregation of such a mixture implies that the OD structure is formed.

{\it The model.}
In this subsection we will assume that the neuron configuration represented by
the neuron densities is given.
It can be any arbitrary configuration, including {\em Salt and Pepper}, {\em
Stripes}, or {\em Patches}.
For the given neuron densities we draw the connections between the cells which:
\begin{itemize}
\item Satisfy the wiring rules ($N_s$ same and $N_o$ other neuron connection
have to be established); 
\item Minimize the total wire length.
\end{itemize}
In the end of the subsection we calculate the wire length for the {\em Salt and
Pepper} configuration.

We consider the mixture of neurons of two types: dominated by the left
and right eyes.  We assume that the neurons are located in the plane.
This assumption is based on the fact that the OD remains constant
in the direction perpendicular to the cortex surface.  The variables
of the problem can therefore be considered functions of the remaining
two coordinates, {\bf r}.

Instead of considering each individual cell we characterize the neuron
configuration by continuous local variables. We define the local
density of neurons dominated by the right eye $n_R({\bf r})$ as the
average density in a square containing sufficiently large number of
cells ($>10$), yet small compared to the typical spatial scales of the
configuration ($\sim 1$mm).  Similarly $n_{L}\left( {\bf r}\right)$ is
the local density of cells dominant by the left eye. Although both
$n_R\left( {\bf r}\right)$ and $n_L\left( {\bf r}\right)$ can vary in
space, the total density of neurons $n_0 \equiv n_R\left( {\bf
r}\right) + n_L\left( {\bf r}\right)$ is a constant, independent of
the position in the cortex.

In our model $n_R\left( {\bf r}\right)$ and $n_L\left( {\bf r}\right)$
completely define the neuron configuration.  For example the {\em Salt
and Pepper} configuration, in which the densities of right-eye and
left-eye neurons are uniform, can be defined as follows
\begin{equation}
\begin{array}{l}
{n_R\left( {\bf r}\right) \equiv \bar{n}_R = f_R n_0,} \\ \\
{n_L\left( {\bf r}\right) \equiv \bar{n}_L = f_L n_0,} \\ \\
{\bar{n}_R + \bar{n}_L = n_0,} 
\end{array}
\label{SaltAndPepper}
\end{equation}
Here $f_R$ is defined as the fraction of the right-eye neurons with
respect to the total number of cells (in general not $1/2$).

Having defined the neuron configuration by fixing the densities $n_R({\bf r})$
and $n_L({\bf r})$ 
we proceed to establishing the connections between cells. 
Two requirements have bo te taken into account. First, we have to satisfy the
wiring rules.
Second, for given densities $n_R({\bf r})$ and $n_L({\bf r})$ the total length
of connections has to be minimum.
Consider a pattern of connections from a neuron dominated, for example, by the
right eye. 
Consider also the region in the cortex it is connected to.
There are in fact two such regions, for the right- and left-eye connections.
We claim that each of these regions is a disc. 
To prove this, notice that if the connections are produced with neurons outside
of this
disk rather than inside the wire length is increased.
This is inconsistent with the requirement of the optimum wiring for a given
configuration.
We denote the radii of these two disks $R_{RR}\left( {\bf r}\right)$ and
$R_{RL}\left( {\bf r}\right)$, implying 
the radii of right-eye neuron at point ${\bf r}$ connection regions to the
right-eye and left-eye cells correspondingly.
Similar quantities can be introduced for the left-eye neurons at point ${\bf
r}$, i. e.
$R_{LR}\left( {\bf r}\right)$ and $R_{LL}\left( {\bf r}\right)$.
We introduce the index notation $i=\{R, L\}$. Then the four radii discussed can
be collapsed 
into one notation $R_{ik}\left( {\bf r}\right)$, standing for the radius of the
connection region 
for the neuron of OD $i$ at point ${\bf r}$
to the cells of OD $k$. The radius can be determined from the wiring rules
($N_s$ and $N_o$ connections to the cells of the same and other OD respectively
have to be 
established):
\begin{equation}
N_{ik} = \int_{\displaystyle \left| {\bf r}-{\bf r}'\right| \le R_{ik}\left(
{\bf r} \right)} d{\bf r}' n_k\left({\bf r}'\right).
\label{EqForRadius}
\end{equation}
Here the elements of matrix $N_{ik}$, $i=\{R, L\}$ are equal to $N_s$ if $i=k$
and $N_o$ otherwise.

It is now possible to determine the total connection length in the cortex ${\cal
L}$. 
To this end we add up the lengths of the connections of individual neurons
$L_{ik}\left({\bf r}\right)$ over the whole area:
\begin{equation}
{\cal L} = \int d{\bf r} n_i\left( {\bf r}\right)\sum_{i, k = R, L}
L_{ik}\left({\bf r}\right),
\label{TotalWireLength}
\end{equation}
where
\begin{equation}
L_{ik} \left({\bf r}\right) 
= \int_{\displaystyle \left| {\bf r}-{\bf r}'\right| \le R_{ik}\left( {\bf r}
\right)} d{\bf r}' n_k\left({\bf r}'\right)
\left| {\bf r} - {\bf r}'\right|.
\label{ConnectionLength}
\end{equation}
The last factor in this expression is the connection length as a function of
separation ${\bf r} - {\bf r}'$ between neurons.
In principle, cost function may not be a linear function of separation. However,
we take it to be linear for the sake of simplicity.
Eqs.~(\ref{EqForRadius}) - (\ref{ConnectionLength}) define our model completely.

Using Eq.~(\ref{TotalWireLength}) we calculate  the wire length for the
homogenous {\em Salt and Pepper}
configuration. To this end we substitute the densities given by
Eq.~(\ref{SaltAndPepper}) into (\ref{EqForRadius})
to find:
\begin{equation}
R^{\rm SP}_{ik} = \sqrt{\frac{N_{ik}}{\pi n_k}}.
\label{R_SandP}
\end{equation}
Then using Eqs.~(\ref{TotalWireLength}) and (\ref{ConnectionLength}) we obtain
\begin{equation}
L_{ik} = \frac 23 R^{\rm SP}_{ik}
\end{equation}
and finally
\begin{equation}
\begin{array}{l}
{\displaystyle {\cal L}^{\rm SP} = \frac {2A}{3} \left[ 
\sqrt{\frac {N_s^3}{\pi}} \left( \sqrt{n_R} + \sqrt{n_L}\right) \right. } \\ \\
{\displaystyle \ \ \ + \left. \sqrt{\frac {N_o^3}{\pi}} \left( \frac
{n_R}{\sqrt{n_L}} + \frac{n_L} {\sqrt{n_R}} \right) 
\right],}
\end{array}
\label{L_SandP}
\end{equation}
where $A$ is the total area of the cortex.

In the next subsection we show that wire length can be reduced with
respect to (\ref{L_SandP}) by introducing a small inhomogeneity into
the neuron densities $n_R$ and $n_L$.  To this end we treat our model
(\ref{EqForRadius}) - (\ref{ConnectionLength}) in the framework of the
perturbation theory.


{\it Instability of the Uniform State Leads to the Formation of
Patterns.}
The purpose of this subsection is to study structures that do not deviate
far from the uniform {\em Salt and Pepper} configuration discussed in
the previous subsection.  Because we have solved the uniform
configuration exactly, the configurations which are not far from it
are also treatable by the perturbation theory analysis i. e. expansion
of the wire length (\ref{TotalWireLength}) in terms of the deviation
of densities of right and left eye neurons from the constant. This
treatment determines which of the inhomogeneous phases ({\em Stripe} or
{\em Patch}) is optimum.  Also, comparison with the numerical results
shows that the perturbation theory results hold even for big differences in
density.
 
We therefore consider a small repositioning of neurons, leading to the
deviation of densities from constant $\delta n\left({\bf r}\right)$.
Because $n_R+n_L=n_0$
\begin{equation}
\begin{array}{l}
{\displaystyle n_R\left({\bf r}\right)=\bar{n}_R + \delta n\left({\bf r}\right),
} \\ \\
{\displaystyle n_L\left({\bf r}\right)=\bar{n}_L - \delta n\left({\bf r}\right).
}
\end{array}
\label{VariationOfDensity}
\end{equation}
As this is only rearrangement
the average of $\delta n\left({\bf r}\right)$ over the entire volume
$\overline{\delta n\left({\bf r}\right)}$
is zero, i. e. the total number of left and right eye neurons is not changed by
the perturbation.
We then substitute these functions into our model (\ref{EqForRadius}) -
(\ref{ConnectionLength}) 
and calculate expansion of the wire length in the Taylor series in $\delta
n\left({\bf r}\right)$. 
It has the form:
\begin{equation}
{\cal L} = {\cal L}^{\rm SP} + {\cal L}^{(1)} + {\cal L}^{(2)} + \ldots,
\end{equation}
where ${\cal L}^{\rm SP}$ is given by Eq.~(\ref{L_SandP}), ${\cal L}^{(1)}
\propto \delta n$,
${\cal L}^{(2)} \propto \delta n^2$ are the first and the second order
corrections to the wire length. 
From the condition $\overline{\delta n\left({\bf r}\right)} = 0$ it follows that
${\cal L}^{(1)}=0$. The second order correction to the wire length is
\begin{equation}
\begin{array}{l}
{\displaystyle {\cal L}^{(2)} = \int d{\bf r} d{\bf r}' \sum_{i,k = R,L} R_{ik}^{\rm
SP}} \\ \\
{\displaystyle \ \ \  \times \left\{ U_{1ik}\left( {\bf r} - {\bf r}'\right)
\delta n_i\left({\bf r}\right) \delta n_k\left({\bf r}'\right) \right. } \\ \\
{ \ \ \ \ \ \ +\left. U_{2ik}\left( {\bf r} - {\bf r}'\right) \delta
n_k\left({\bf r}\right) \delta n_k\left({\bf r}'\right) \right\} .}
\end{array}
\label{SecondOrder}
\end{equation}
Here $\delta n_i\left({\bf r}\right)$ is the perturbation of density of neurons
of $i$-th dominance
($\delta n_R = \delta n$, $\delta n_L = -\delta n$,) and
\begin{equation}
U_{1ik}\left( {\bf r}\right) = \theta \left( R_{ik}^{\rm SP} - \left| {\bf r}
\right| \right) 
\left(  \frac { \left| {\bf r} \right| } { R_{ik}^{\rm SP} } - 1\right),
\label{FirstKernel}
\end{equation}
\begin{equation}
\begin{array}{l}
{\displaystyle U_{2ik}\left( {\bf r}\right) = \frac {1}{ 4\pi \left( R_{ik}^{\rm
SP} \right) ^ 2 } } \\ \\
{\displaystyle \ \ \ \times \int d {\bf r}'' 
\theta \left( R_{ik}^{\rm SP} - \left| {\bf r} - {\bf r}''\right| \right) 
\theta \left( R_{ik}^{\rm SP} - \left| {\bf r}'' \right| \right) }
\end{array}
\label{SecondKernel},
\end{equation}
where $\theta (x) = 1$, if $x \ge 0$, and $\theta (x) = 0$, if $x < 0$.
Because $U_{2ik}$ has the geometrical  interpretation of the overlap between two
disks:
\begin{equation}
\begin{array}{l}
{\displaystyle  U_{2ik}\left( {\bf r}\right) = \left[ \frac {1}{ 2\pi } \arccos
\left( \frac{ \left| {\bf r} \right| }{ 2R_{ik}^{\rm SP} } \right) \right.} \\
\\
{\displaystyle \ \ \  \left. - \frac{ \left| {\bf r} \right| }{ 4\pi R_{ik}^{\rm
SP} } 
\sqrt { 1 - \left( \frac { \left| {\bf r} \right| } {2R^{\rm SP}_{ik}} \right)^2
}
\right] \theta \left( 2R_{ik}^{\rm SP} - \left| {\bf r} \right| \right) .}
\end{array}
\label{SecondKernelExact}
\end{equation}
Using Eq.~(\ref{VariationOfDensity}) we express the second order correction to
the wire length (\ref{SecondOrder})
as a pairwise density-density interaction
\begin{equation}
{\cal L}^{(2)} = \int d{\bf r} d{\bf r}' \delta n\left({\bf r}\right) 
{\cal U} \left( {\bf r} - {\bf r}'\right)  \delta n\left({\bf r}'\right),
\label{SecondOrderHam}
\end{equation}
where the ``interaction potential'' ${\cal U} \left( {\bf r} \right)$ is given
by
\begin{equation}
{\cal U} \left( {\bf r}\right) = {\cal U}_1\left( {\bf r}\right) + {\cal
U}_2\left( {\bf r}\right),
\end{equation}
\begin{equation}
{\cal U}_1 = U_{1RR} + U_{1LL} - U_{1RL} - U_{1LR},
\label{U1}
\end{equation}
\begin{equation}
{\cal U}_2 = U_{2RR} + U_{2LL} + U_{2RL} + U_{2LR}.
\label{U2}
\end{equation}

We notice that expression \ref{SecondOrderHam} is similar to the Hamiltonian used
by Cowan and Friedman, 1991, which corresponds to Swindale, 1980,
learning rules. The advantage of our approach is that we derive this expression
from a single principle without assuming a particular form of ``interaction
potential''. In addition, we go on to solve this expression analytically. This 
allows us to map out a phase diagram which relates the ocular dominance pattern to
biologically measurable connection rules without appealing to ``Mexican hat''
connection weights. 

We convert Eq.(\ref{SecondOrderHam}) into Fourier space using the property of a
convolution
\begin{equation}
\displaystyle
{\cal L}^{(2)} = \int \frac{d{\bf q}}{(2\pi)^2} 
\tilde{{\cal U}} \left( {\bf q} \right)
\left|\delta \tilde{n}\left({\bf q}\right)\right|^2,
\label{SecondOrderHamQ}
\end{equation}
where $\tilde{{\cal U}} \left( {\bf q} \right)$ and $\delta \tilde{n}\left({\bf
q}\right)$
are Fourier transforms of the ``interaction potential'' and the perturbation of
density 
respectively. The Fourier transform of a function $f({\bf r})$ is defined as 
$\tilde{f}({\bf q})=\int d{\bf r} f({\bf r}) \exp(-i{\bf q}{\bf r}),$ where
$i=\sqrt{-1}$ is
the imaginary unity. Eq.~(\ref{SecondOrderHamQ}) is the central result of this
subsection.

Function $\tilde{{\cal U}} \left( {\bf q} \right)$ determines the changes in the
total wire length
due to the deviation of the neuron density from constant. 
For example, if the perturbation of density has the form of plane wave ($\delta
n = a\cos({\bf q}_0{\bf r})$),
the change in the total wire length is proportional to  $\tilde{{\cal U}} \left(
{\bf q}_0 \right)a^2$.
Thus if $\tilde{{\cal U}} \left( {\bf q} \right)$ is negative at certain $q_0$, 
such a perturbation {\em decreases} the total wire length. 
It is therefore advantageous from the point of view of wire length economy to
create a 
perturbation of density at this wave vector. 
In this case the uniform {\em Salt and Pepper} configuration ($\delta n = 0$)
is unstable with respect to the formation of the OD patterns.
Hence, negative function $\tilde{{\cal U}} \left( {\bf q} \right)$ indicates the
formation an OD pattern.

We therefore analyze the conditions at which the function has negative values.
Two statements can be made in this respect.
First, assume that $N_s=N_o$. Then $\tilde{{\cal U}} \left( {\bf q} \right)$ is
never negative.
This implies that {\em Salt and Pepper} is optimum if $N_s=N_o$.
Second, consider $\tilde{{\cal U}} \left( {\bf q} \right)$ at $f_R=f_L=1/2$ and
arbitrary $N_s \neq N_o$.
In this case $\tilde{{\cal U}} \left( {\bf q} \right)$ always has negative
values. 
This means that on the line of equal right-left eye occupancy $f_R=f_L=1/2$ the
OD 
patterns are {\em always} optimum, except for the point $N_s=N_o$.
We do not give the proofs of these properties due to the space limitations. 
 
%
%
\begin{figure}
\centerline{
\psfig{file=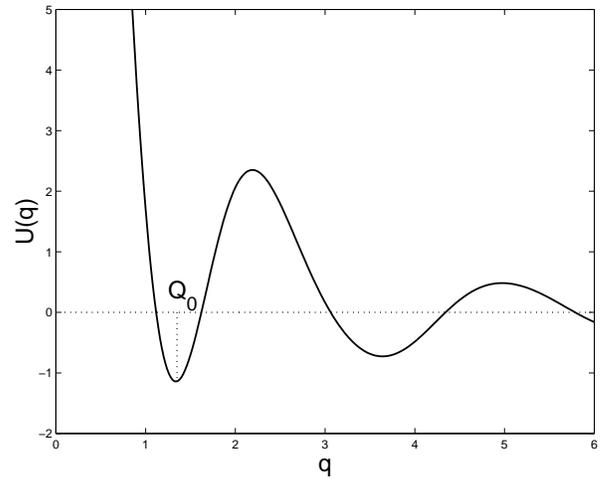,width=3.1in}
}
\vspace{0.1in} 
\setlength{\columnwidth}{3.4in}
\caption{Function \protect{$\tilde{{\cal U}} \left( {\bf q}
\right)$}
calculated numerically for \protect{$N_s=10$, $N_o=7$, $f_R=1/2$, and $n_0=1$.}
The value of wave vector corresponding to the most negative value of the
function
is denoted \protect{$Q_0$}.
\label{fig30}
}
\vspace{0.1in}
\end{figure}


To illustrate these properties we show an example of $\tilde{{\cal U}} \left(
{\bf q} \right)$ 
in Fig.~\ref{fig30}. The function obviously has negative values, signifying
instability and 
an OD pattern formation. The instability
is strongest at the wave vector corresponding to the most negative value of 
$\tilde{{\cal U}} \left( {\bf q} \right)$. Indeed, creating the structure at
this 
wave vector reduces the total wire length most effectively. We predict therefore
the spatial period of the OD pattern. 
For the case $N_s\approx N_o$ shown in Fig.~\ref{fig30} ($N_s=10$ and $N_o=9$)
function $\tilde{{\cal U}} \left( {\bf q} \right)$ reaches the most negative
value at
\begin{equation}
Q_0 \approx \frac 3{R^{\rm SP}_{RR}} \approx \frac 3{R^{\rm SP}_{RL}}
\label{OptimumFrequency}
\end{equation}
The spatial period of the OD pattern is therefore
\begin{equation}
\Lambda = \frac {2\pi}{Q_0} \approx 2R^{\rm SP}_{RR} \approx 2R^{\rm SP}_{RL}.
\end{equation}
In other words it is approximately equal to the diameter of the disc of
connections.

\begin{figure}
\centerline{
\vspace{0in}
\psfig{file=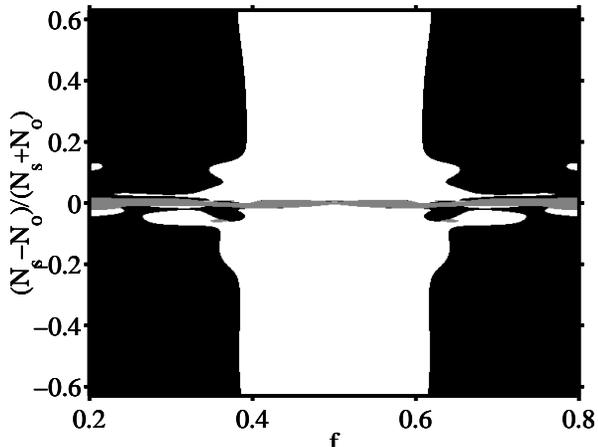,width=3.1in}
}
\vspace{0in} 
\setlength{\columnwidth}{3.4in}
\caption{Ocular dominance phase diagram calculated in 
perturbation theory. Range of the {\em Stripe} phase optimality is shown in
black, {\em Patch} phase - in white, {\em Salt and Pepper} - in grey.
\label{fig40}
}
\vspace{-0.1in}
\end{figure}

{\it Competition between the {\it Stripe} and {\it Patch} phases}
Next we use the perturbation theory to calculate approximately the
wire lengths of different OD structures. Because the structures 
are periodic the integral in Eq.~(\ref{SecondOrderHamQ}) can be reduced to the
sum over the reciprocal lattice vectors ${\bf Q}$:
\begin{equation}
\displaystyle
{\cal L}^{(2)} = \frac{1}{A} \sum_{{\bf Q} \neq 0} \tilde{{\cal U}} \left( {\bf
Q} \right)
\left|\delta \tilde{n}\left({\bf Q}\right)\right|^2,
\label{SumOverRecVect}
\end{equation}
where $A$ is the total area of the system. 
Different OD structures have different sets of ${\bf Q}$ and $\delta
\tilde{n}\left({\bf Q}\right)$.
For example, for {\em Stripes} $Q_x = 2\pi n/\Lambda$, $Q_y=0$, where $n=\pm 1,
\pm 2, \ldots$ and $\Lambda$ is the
spatial period of the structure. The Fourier transform of density
\begin{equation}
\delta \tilde{n}_{\rm Stripes}\left({\bf Q}\right) = 
\frac {2A}{\Lambda\left|{\bf Q}\right|} \sin\left(\frac{f_R\left|{\bf
Q}\right|\Lambda}{2}\right).
\label{LammelarDensity}
\end{equation}
For the triangular lattice of {\em Patches} $Q_x = Q_0(l\sqrt{3}/2)$, $Q_y =
Q_0(k+l/2)$, with 
$l,k = \pm 1, \pm 2, \ldots$ and $Q_0=4\pi/\Lambda\sqrt{3}$,
where $\Lambda$ is the lattice spacing. The Fourier transform of density
\begin{equation}
\delta \tilde{n}_{\rm Patches}\left({\bf Q}\right) =
\frac{2A}{\Lambda\left|{\bf Q}\right|} \sqrt{\frac{2\pi f_R}{\sqrt{3}}} 
\sin\left( \left|{\bf Q}\right|\Lambda \sqrt{\frac{f_R\sqrt{3}}{2\pi}} \right).
\label{PatchesDensity}
\end{equation}
Based on Eqs.~(\ref{SumOverRecVect})-(\ref{PatchesDensity}) we compare different
OD structures and 
generate the phase diagram similar to one given in the introduction (see
Fig.~\ref{fig40}).

\vspace{-0.2in}
%
%

Figures \ref{ph_diagram} and \ref{fig40} have many similar features. First,
the diagram is symmetric with respect to the line $f_R=1/2$. This is a
consequence of the 
left-right eye symmetry of the general wire length functional
(\ref{TotalWireLength})
inherited by the second order functional (\ref{SecondOrderHamQ}).
The reason for the existence of such a symmetry is interchangeability of left and
right eyes
inherent to this model.
If, for instance, in a given configuration one relabels left-eye neurons into
the right-eye ones 
and wise versa, the wire length does not change.

Second, the {\em Salt and Pepper} phase occupies a stripe around the line
$N_s=N_o$. 
The width of this line is given by $|N_s-N_o| < 0.01N_s$. This
is the result of the above-mentioned stability of {\em Salt and Pepper} on the
line $N_s=N_o$. 
As it is shown by the diagram the stability extends into some region around this
line.

Third, there is a transition between {\em Stripes} and {\em Patches} at $f_R
\approx 0.4$ and 
$0.6$. The region on the diagram corresponding to $0.4 < f_R < 0.6$ is almost
completely occupied 
by the {\em Stripes} while the rest of the diagram ($f_R<0.4$ and $f_R>0.6$)
by the {\em Patches}.
We explain this in the framework of the perturbation theory. 
The main contribution to Eq.~(\ref{SumOverRecVect}) comes from the terms with
the smallest $|{\bf Q}|$.
This happens because both $\tilde{{\cal U}} \left( {\bf Q} \right)$ 
and $\tilde{n}\left({\bf Q}\right)$ decay very fast with the increase of $|{\bf
Q}|$.
{\em Stripes} and {\em Patches} can approximately be compared using only the 
terms with the smallest $|{\bf Q}| \equiv Q_0$. 
The two solutions have equal wire length if
\begin{equation}
2\tilde{{\cal U}} \left( Q_0 \right)
\left|\delta \tilde{n}_{\rm Stripes}\left(Q_0 \right)\right|^2
=
6\tilde{{\cal U}} \left( Q_0 \right)
\left|\delta \tilde{n}_{\rm Patches }\left(Q_0\right)\right|^2,
\end{equation}
where the factors $2$ and $6$ are the numbers of the smallest wave length
harmonics
in the {\em Stripe} and {\em Patch} phases respectively. 
Using Eqs.~(\ref{LammelarDensity}) and (\ref{PatchesDensity})
to solve the latter equation for $f_R$ we obtain numerically for
the filling factor of the transition $f_R \approx 0.4$. Due to the mentioned
left-right eye symmetry of the model similar transition occurs at $f_R = 1 -
0.4 = 0.6$.

We would like to notice finally that comparison of the perturbation
theory to exact calculations shows that the former works well even if the
deviation of the 
density from constant is not small ($\sim 0.5n_0$). Such a comparison shows that 
$\left( \left[{\cal L} - {\cal L^{\rm SP}}\right] - {\cal L}^{(2)}\right)/{\cal
L}^{(2)} < 5\%$.
In addition the perturbation theory provides a framework to understand numerous
qualitative 
features of the phase diagram discussed above.

Von der Malsburg, 1979, has surmised that there is a phase transition 
between {\it Patches} and {\it Stripes} driven by the cost of the left/right 
eye boundary. However he did not address different numbers of connections with 
same vs. other-eye neurons and made several different assumptions (e.g. fixing 
the periodicity of the pattern). 
Thus our results offer a more complete description of the OD patterns while 
relying only on one principle - wire length minimization.


\section{Discussion}

\subsection{Comparison with experiment}

This theory relates functional requirements on a neural circuit to its
structural properties. In particular, the phase diagram relates the
relative fractions of neurons, $f_L$, and of connections, $N_s/N_o$, to
the appearance of the OD pattern. Ideally, this theory
could be tested by measuring these numbers experimentally and
comparing the observed OD pattern to the one predicted by the phase
diagram. However, we could not find data on the ratio
$N_s/N_o$ and can only surmise that it is greater than one.

We can partially test the theory by using the predictions of the phase
diagram which are independent of the ratio $N_s/N_o$. Fig.\ref{phases} shows
that the transition from the {\em Stripe} and the {\em L-Patch} ({\em
R-Patch}) phase takes place when $f_L\approx 0.4$ ($f_R\approx 0.4$)
for a wide range of $N_s/N_o$. This number can be compared with the
experimentally derived value of $f_L$ which is found from the relative
area occupied by left-eye dominated neurons. The prediction that the
{\em Patch} phase becomes optimal when one eye dominates is, indeed, non-trivial
because there may be a system of alternating wide and narrow monocular
stripes instead.

\begin{figure}
\centerline{
\psfig{file=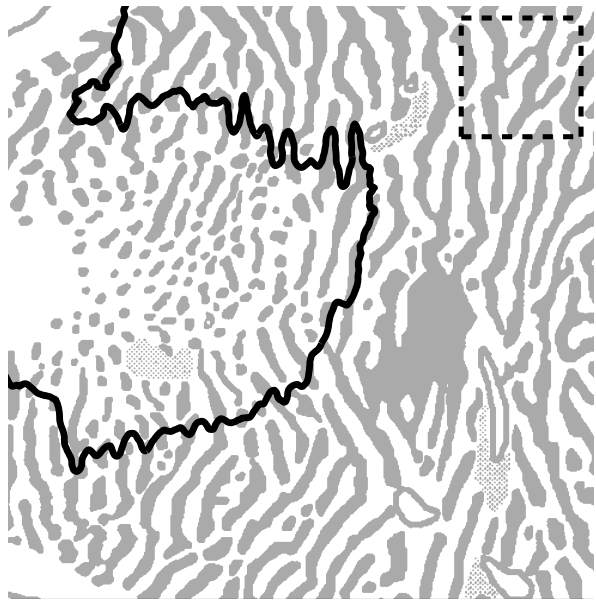,width=3.1in}
}
\vspace{0.1in} 
\setlength{\columnwidth}{3.4in}
\caption{Transition between the {\em Stripe}
and {\em Patch} phases occurs at theoretically predicted value
$f_L$. Shown is a fragment of the macaque ocular dominance pattern
from Horton and Hocking, 1996. Neurons dominated by the
left eye are grey and neurons dominated by the right eye are white. 
Black contours correspond to the
value {\protect $f_L=0.4$} averaged over a window equal to the one
shown (dashed lines). Transition from {\em Stripe} to {\em Patch} 
phase visually coincides with the black contour.
\label{macaque}
}
\vspace{-0.2in}
\end{figure}

We test this prediction on the data from macaque and {\em Cebus}
monkey. The relative area occupied by the left/right eye depends on
the location in V1. In para-foveal part of V1 both eyes are represented
equally, $f_R\approx 0.5$. In agreement with the phase diagram, the OD
pattern consists of stripes. Farther from the fovea contralateral eye
becomes dominant. The OD pattern becomes patchy there, just as
expected from the phase diagram. We verify the location of the
transition by using the following algorithm. We find $f_L$ for each
point of the pattern by calculating the relative area occupied by the
left/right regions in a window centered on that point and a few OD
periods wide (dashed lines in Fig.\ref{macaque}). Then we draw a contour 
corresponding to $f_L=0.4$,
Fig.\ref{macaque}. Next we check visually whether the location of this
contour is close to the transition from {\em Stripes} to {\em
Patches}.  Indeed, the large black contour in Fig.\ref{macaque} coincides
with the transition indicating good agreement.

In {\em Cebus} monkey the OD pattern has a similar transition~\cite{Cebus}.
For monkey CO6L from Rosa et al., 1992, we determine visually that 
along the horizontal meridian the transition occurs at the eccentricity of 
$20-40\deg$. According to the plot of the relative representations given
in  Rosa et al., 1992, $f_L$ changes in the range $0.32-0.42$ at these
eccentricities. Our prediction of $f_R=0.4$ falls into this interval. For 
the upper $45$ degree meridian of the same monkey
the transition occurs at the eccentricity of $30-40$ degrees or at filling
fractions $0.33-0.43$. Again, the predicted value belongs to this
interval.
We conclude that this data agrees with our predictions although a more
precise measurement would be helpful.

In cats the OD patterns resemble {\em Patches}. In this case our
theory implies that one eye should dominate. In fact, Shatz and
Stryker, 1978, reported that the filling fraction of the
contralateral eye in cat V1 is greater than $0.5$. This may explain
the existence of {\em Patches} in cat V1. However, other
authors~\cite{Anderson88} claimed that both eyes are represented almost
equally. More precise measurements of the ocular dominance are
needed to make a conclusive judgment.

%
%

\subsection{Further Development of the Theory}

Next, we elaborate on several simplifying assumptions made in the paper. 
Although these assumptions should not affect our conclusions
significantly, 
they are worth further exploration.

First, the transition between {\em Stripes} and {\em Patches} may be
more complex than discussed. We considered only two candidate phases:
{\em Stripes} and a triangular lattice of circular {\em Patches}. It
is possible that some intermediate phases become optimal near the
transition. For example, Fig.\ref{macaque} hints that parallel
chains of elongated {\em Patches} may give more efficient wiring. This
would slightly modify our phase diagram.

Second, we based our theory on looking for an optimal layout of
neurons which minimizes total wire length. The considered structures
are, therefore, regular and periodic. However, developmental noise may
lead to fluctuations in the OD pattern which reduce
slightly its wiring efficiency.  Although actual OD patterns contain
such fluctuations we do not know whether these are due to suboptimal
wiring or variations in the wiring rules from point to point.

Different phases may have different stability in respect to
noise. Judging from the data, the {\em Stripe} phase holds up well on
the scale of a few periods. The {\em Patch} phase, however, does not
show a regular triangular lattice. We think that this is because of a
relatively small difference in wire length between the triangular and
the square lattice of {\em Patches}.  (It is about $0.5\%$ of the
total wire length, compared to $2\%$ difference between {\em Stripes}
and {\em Patches} for the upper left part of the phase diagram.)

Third, we ignored the presence of numerous other cortical maps, such
as orientation selectivity and retinotopy of receptive fields. This
follows from the assumption that the mutual interaction (or coupling)
between different maps is weak.  For example, as mentioned in the
Introduction, our motivation to neglect retinotopy comes from the
magnitude of the receptive field scatter exceeding the width of ocular
dominance stripes.  The best justification for ignoring the coupling
between the maps comes from the robustness of the observed OD patterns
and the good agreement of our theory with experiment.

Our theory can be expanded to address the interaction between
different maps.  Variables of other maps can enter the expression for
the total wire length, Eq.\ref{TotalWireLength}, through additional
values of indices $i$ and $k$, which so far reflect ocular
dominance. Moreover, these indices can become continuous variables if
the sums (\ref{TotalWireLength}) are replaced by integrals. This would
be appropriate for including interactions with retinotopic and
orientational selectivity maps.

Fourth, we applied our theory to the OD patterns as the best studied
structure.  Since our model is based on minimal assumptions, it can be
applied to other binary structures such as cytochrome oxidase blobs.

In conclusion, we explained the OD patterns in mammalian V1 by minimizing 
wire length given general functional considerations. Good agreement with
experiment lends strong support to the notion that OD structures are 
adaptations to reduce intra-cortical wiring.


\end{multicols}

\begin{thebibliography}{99}
\small
\bibitem[Ahmed et al., 1994]{Ahmed} Ahmed B, Anderson JC, Martin KA, \& Nelson JC  
(1994) Polyneuronal innervation of spiny stellate neurons in cat visual cortex. 
J Comp Neurol. 341: 39-49. 

\bibitem[Allman and Kaas, 1974]{AK} Allman JM \& Kaas JH (1974) The organization 
of the second visual area (V II) in the owl monkey: a second order transformation 
of the visual hemifield. Brain Res 76: 247-65.

\bibitem[Anderson et al., 1988]{Anderson88} Anderson PA, Olavarria J, \& Van 
Sluyters RC (1988) The overall pattern of ocular dominance band in cat visual 
cortex. J of Neurosci 8:2183-2200.

\bibitem[Cajal, 1995]{Cajal} Cajal SRy (1995) Histology of the nervous system 
1-805 (Oxford University Press, New-York).

\bibitem[Cherniak, 1992]{Chern} Cherniak C (1992) Local optimization of neuron 
arbors. Biol Cybern 66: 503-10.

\bibitem[Chklovskii and Stevens, 1999]{CS} Chklovskii DB \& Stevens CF  (1999)
Wiring the brain optimally. Submitted Nature Neuroscience.

\bibitem[Cowan and Friedman, 1991]{Cowan} Cowan JD \& Friedman AE (1991) Simple 
spin models for the development of ocular dominance columns and iso-orientation 
patches. {\it Advances in Neural Information Processing} {\bf 3}, Eds Lippmann R,
Moody J \& Touretzky D, Morgan Kaufmann, 26-31.

\bibitem[Cowey, 1979]{Cowey} Cowey A (1979) Cortical maps and visual perception: 
the Grindley Memorial Lecture. Q J Exp Psychol 31: 1-17.

\bibitem[Erwin et al., 1995]{Erwin} Erwin E, Obermayer K \& Schulten K  (1995) 
Models of orientation and ocular dominance columns in the visual cortex: 
a critical comparison. Neural Comput 7: 425-68.

\bibitem[Horton and Hocking, 1996]{Macaque} Horton JC \& Hocking DR (1996) 
Intrinsic variability of ocular dominance column periodicity in normal macaque 
monkeys. J Neurosci 16: 7228-39.

\bibitem[Hubel and Wiesel, 1974]{HubWies74} Hubel DH \& Wiesel TN (1974)
Uniformity of monkey striate cortex: a parallel relationship between field size, 
scatter, and magnification factor. J Comp Neurol 158: 295-305.

\bibitem[LeVay and Gilbert, 1976]{LVG} LeVay S \& Gilbert CD (1976) Laminar 
patterns of geniculocortical projection in the cat. Brain Res 113: 1-19.

\bibitem[Malsburg, 1979]{Malsburg} von der Malsburg C (1979) Development of 
ocularity domains and growth behaviour of axon terminals. Biol Cybern 32: 49-62.

\bibitem[Mitchison, 1991]{Mitch} Mitchison G  (1991) Neuronal branching patterns 
and the economy of cortical wiring. Proc R Soc Lond B Biol Sci 245: 151-8.

\bibitem[Mountcastle, 1957]{Mou} Mountcastle VB  (1957) J Neurophysiol 20: 408-434.

\bibitem[Peters and Payne, 1993]{PP} Peters A \& Payne BR (1993) A numerical 
analysis of the geniculocortical input to striate cortex in the monkey. 
Cereb Cortex 4: 215-29. 

\bibitem[Rosa et al., 1992]{Cebus} Rosa MG Gattass R, Fiorani M Jr \& Soares JG
(1992) Laminar, columnar and topographic aspects of ocular dominance in the 
primary visual cortex of Cebus monkeys. Exp Brain Res 88: 249-64.

\bibitem[Shatz and Stryker, 1978]{shatz78} Shatz CJ \& Stryker MP (1978) Ocular 
dominance in layer IV of the cat's visual cortex and the effects of monocular 
deprivation. J Physiol (Lond) 281: 267-83 (1978).

\bibitem[Swindale, 1980]{Swindale} Swindale NV (1980) A model for the formation 
of ocular dominance stripes. Proc R Soc Lond B Biol Sci 208: 243-64.

\bibitem[Swindale, 1996]{SwinRev} Swindale NV (1996) The development of topography 
in the visual cortex: a review of models. Network: Computation in Neural 
Systems 7: 161-247.

\bibitem[Wiesel and Hubel, 1974]{WHL} Wiesel TN, Hubel DH \& Lam DM (1974) 
Autoradiographic demonstration of ocular-dominance columns in the monkey striate 
cortex by means of transneuronal transport. Brain Res 79: 273-9.

\bibitem[Young, 1992]{Young} Young MP (1992) Objective analysis of the 
topological organization of the primate cortical visual system. Nature 358: 152-5.


\end{thebibliography}
\end{document}